\documentstyle[12pt]{article}
\newcounter{multieqs}

\newcommand{\bq}{\begin{equation}}
\newcommand{\fq}{\end{equation}}
\newcommand{\bqr}{\begin{eqnarray}}
\newcommand{\fqr}{\end{eqnarray}}
\newcommand{\non}{\nonumber \\}

\newcommand{\rf}[1]{(\ref{#1})}

\def\plb#1#2#3{Phys. Lett. {\bf{#1B}} (#2) #3}

\def\atmp#1#2#3{Adv. Theor. Mat. Phys. {\bf #1} (#2) #3}
\def\del{\delta}    
        
       \def\lam{\lambda}
    
 \def\vphi{\varphi}

 \def\cN{{\cal N}}

\def\pa{\partial}

\def\ove#1{\frac{1}{#1}}

\def\bk{{\vec{k}}}
\begin{document}

\thispagestyle{empty}

\marginparwidth = .5in

% The following command moves parenthetical comment.
\marginparsep = 1.2in

\begin{flushright}
\begin{tabular}{l}
ANL-HEP-PR-99-77 \\ 
\\

hep-th/9907015
\end{tabular}
\end{flushright}

\vspace{18mm}
\begin{center}

{\bf The Intermediate Coupling Regime in the AdS/CFT Correspondence}

\vspace{18mm}

{Gordon Chalmers}\footnote{E-mail address: chalmers@pcl9.hep.anl.gov}
\\ [10mm]
{\em Argonne National Laboratory \\
High Energy Physics Division \\
9700 South Cass Avenue \\
Argonne, IL  60439-4815 } \\

\vspace{20mm}

{\bf Abstract}

\end{center}

The correspondence between the 't Hooft limit of $N=4$ super 
Yang-Mills theory and tree-level IIB superstring theory on 
AdS${}_5\times S^5$ in a Ramond-Ramond background at values 
of $\lambda=g^2 N$ ranging from infinity to zero is examined 
in the context of unitarity.  A squaring relation for the 
imaginary part of the holographic scattering of identical 
string fields in the two-particle channels is found, 
and a mismatch between weak and strong 't Hooft coupling is pointed 
out within the correspondence.  Several interpretations and 
implications are proposed.

\vfill

\setcounter{page}{0}
\newpage
\setcounter{footnote}{0}

\baselineskip=16pt

\section{Introduction} 

Recently there has been much interest in understanding the 
't Hooft limit of $N=4$ super Yang-Mills theory at strong coupling via the 
conjectured correspondence with IIB superstring theory 
compactified on $AdS_5\times S^5$ in a non-vanishing Ramond-Ramond 
background \cite{mal,gub,wit,rev}.  Despite the developments in 
technology useful in computing correlation functions at 
large $\lambda$ through the classical holographic supergravity 
description, there has been little work on interpolating  
correlation functions beyond three-point between the two regimes.  As 
the four-point functions are known not to correspond to the free-field 
limit at large coupling, this would involve resumming an infinite set 
of planar Feynman diagrams and comparing with the scattering element 
derived from the infinite number of string exchanges in tree-level 
string theory.  

Both these calculations appear hard in that summing an infinite 
number of Feynman diagrams seems intractable, while the techniques 
of perturbative string theory on $AdS_5\times S^5$ in a 
Ramond-Ramond background have not been fully developed; progress 
along the direction in \cite{RRback,boershat} would be needed.  The general 
four-point function at string tree level in the anti-de Sitter 
(RR) background is a function $F(\lambda, k_i)$ of the 't Hooft 
coupling $\lambda$.  At large $N$ and finite $\lambda=g^2 N$,  
scattering elements obey a factorization 
condition inherited from the classical field equations \cite{cs2}; 
for those between identical fields the factorization condition 
may be used to obtain a non-negativity condition on the imaginary part 
of four-point functions in the two-particle channels.  
On the other hand, the result at $\lambda=0$ in $N=4$ super Yang-Mills 
theory for these dual correlators is easily computable at 
four-point (in $x$-space as well as in $k$-space).  
We shall combine these two calculations in the different limits 
to probe the conjectured correspondence at finite 't Hooft coupling 
and at large $N$.   

Extending the weakly coupled string theory calculation to the 
regime of strong coupling may invalidate the classical approximation 
(the anti-de Sitter space is preserved under quantum corrections to 
leading order corrections in the IIB effective action given in \cite{cor}).  
However, it is of interest to find potential properties of IIB 
holographic string theory on AdS that may lift to the dual boundary 
field theory in the intermediate coupling regime (i.e. $\lambda \sim 
1$).  There are several forms of the conjectured correspondence between 
IIB string theory on AdS${}_5\times S^5$ in a background RR field 
and $N=4$ super Yang-Mills theory; we note beforehand that further 
non-perturbative processes are in accord with summing 
over different asymptotically AdS spacetimes.  Our intention here is 
to examine the holographic scattering in the spirit of the Gross-Mende 
effect \cite{gm} (limiting to a tensionless $R^2/\alpha'\rightarrow 0$ string).

Rather than computing the scattering element on $AdS_5\times S^5$ 
in a constant Ramond-Ramond background,  
we shall use a unitarity argument to find a positivity 
condition on the imaginary part in the two-particle channel for 
particular correlator examples where all of the external fields are of the 
same type.  We assume that unitarity is sensible in the bulk theory, and 
we use the $i\varepsilon$ prescription provided in \cite{cs2}, following from 
the naive Wick rotation of the Euclidean geometry. (See \cite{boxun} 
and \cite{boxcom} for a detailed account of the $i\varepsilon$ in  
critical and compactified four-point string amplitudes.)  Unitarity 
has been examined 
in detail in \cite{vkl,cs2, vgl}.  In the following, we define 
$s=s_{12}$, $t=s_{23}$ and $u=s_{13}$; then $s+t+u=\sum k_j^2$.  
Holding $k_j^2\leq 0$ together 
with $t,u<0$ and $s>0$ we extract the imaginary parts in the 
s-channel by complex conjugation \cite{cs2}.  In conventional 
perturbative string theory on a curved space at any finite value of 
the $s$ value only a finite number of Kaluza-Klein and string 
states may contribute to the spectral density, and hence to 
the imaginary parts at finite $s$.  In the holographic formulation an infinite 
number of string modes contribute to the imaginary parts.  The 
analysis performed here relies only on the existence of field 
equations for the modes of the string on the compactified space, 
thus avoiding the technical complications associated with a path 
integral quantization.  
We further assume that a consistent truncation 
exists for the field equations coming from the compactified string 
theory; this has been shown for the KK compactification of eleven dimensional 
supergravity in an $AdS_4\times S^7$ background \cite{dewitnicolai} and 
recently in the 
$AdS_7\times S^4$ context \cite{truncations}; although 
this has not been shown technically in the case at hand, we shall  
assume that it exists. 

The cut in the $s$-channel, if we choose 
the external lines to be of the same particle type, is of the 
form 
\bq  
{\rm Im}_s A_{ffff}(k_j) = \sum_i M_i(k_1,k_2;k_1+k_2) 
 M_i(k_3,k_4;k_3+k_4) \ ,
\label{factcond} 
\fq 
after integrating over the bulk (fifth) holographic coordinate 
perpindicular to the boundary.  The sum over $i$ extends over all 
contributing intermediate states in the particular holographic 
four-point function.  We now further examine the scalar 
contributions.  The factorization condition follows 
from decomposing the bulk-bulk propagator through the relation 
\bq  
G(x,y) = - \sum_n {\phi_n^*(x) \phi_n(y) \over \lambda_n^2 -i\varepsilon} \ ,
\label{complete}
\fq 
where $\phi_n(x)$ span a complete set of eigenfunction solutions of the 
kinetic operator (with appropriate boundary conditions) for the particular 
field we are considering.  For example, the propagator for massive scalar 
fields has the form after a Wick rotation (in Minkowski anti-de Sitter 
Poincare coordinates $ds^2=1/x_0^2 (dx_0^2+d{\vec x}^2)$ where we 
have the boundary four-dimensional metric given by $u\cdot v = u_0 v_0 - 
{\vec u}\cdot {\vec v}$),  
\bq
G(x,y) = -i \int_0^{\infty} d\lam \,
 \lam \int \frac{d^dk}{(2\pi)^d}
 \frac{\vphi^{\ast}_{\lam}(x)\vphi_{\lam}(y)}{(\lam^2-\bk^2 -i\varepsilon)} \ ,
\label{csBBprop}
\fq
and obeys
\bq
\hat{K} G(x,y) = - i x_0^{d+1}\del^{d+1}(x-y) =
-i \frac{\del^{d+1}(x-y)}{\sqrt{g}} \ , 
\fq 
where
\bq  
\hat{K}= - \left(\ove{\sqrt{g}}\left(\pa_{\mu} \sqrt{g} g^{\mu\nu} 
\pa_{\nu}\right) -m^2 \right) \ .
\fq 
The eigenfunctions $\vphi_\lambda (y)$ obeying the correct Dirichlet 
boundary conditions at $x_0=0$ are 
\bqr
\vphi_{\lam}(x) &=& x_0^{d/2} e^{i\vec{k} \cdot \vec{x}}J_{\nu}(\lambda x_0),
 ~~\vec{k} \cdot \vec{x} \equiv \sum_{i=1}^d k_ix^i
\non
\hat{K} \vphi_{\lam}(x) &=& -(\lam^2-\vec{k}^2)x_0^2 \vphi_{\lam}(x) 
\label{ref}
\ ,
\fqr
where $\nu =\sqrt{m^2+d^2/4} >0$. In Poincare coordinates $\vphi_{\lam}(x)$ 
are labelled by the four-vector $\vec{k}$, the conserved momentum along 
the boundary, and a continous eigenvalue $\lam$.
The $i\varepsilon$ prescription is chosen to agree with that on the 
boundary $N=4$ super Yang-Mills theory via a Wick rotation.  
The result for the correlation function will be proportional to 
a factor of $i$, both in the string theory and in the boundary 
theory; as we are exploring the analytic properties of the 
correlation function we shall drop the factor of $i$ in 
\rf{csBBprop} as well as in the four-point function.  

Further modifications of the Greens functions 
in Minkowski space via adding normalizable zero modes to them \cite{vkl, vklt} 
will not alter the factorization of the general four-point function, but 
rather change the functional form of $M_i$ in \rf{factcond} by the 
addition of terms to the propagators.  Although we have reproduced 
the explicit form of the massive propagator in \rf{csBBprop}, the 
factorization condition found from the complete set of states in 
\rf{complete} is expected to hold for general fields with spin as does 
in perturbative string theory.   

Holding $k^2 <0$ in \rf{csBBprop} extracts the imaginary part within 
the $k$-integration (via ${\rm Im} (\lambda^2-\vec{k}^2-i\varepsilon)^{-1} 
 = -\pi \delta^{(d)}(\lambda^2-\vec{k}^2)$) for 
the propagating modes in the interior of the anti-de Sitter spacetime.  
This imaginary part reproduces the factorization formula in 
\rf{factcond}. 
 
Note that, for an arbitrarily massive mode, the imaginary part 
picks up a contribution from a small negative value of $k^2$, as 
is clear from \rf{csBBprop} and \rf{ref}.  Each term in the sum in \rf{factcond} 
gives rise to a power series expansion in $\lambda$, from expanding the 
mass dependence in the integrals over the bulk-boundary-boundary three-point 
functions (the argument of the Bessel function depends on the mass).  The 
fact that an infinite number of states contribute to the unitarity cut 
at finite $s$ is a holographic feature differing from  
usual perturbative string theory; the latter gives contributions in accord 
with particle thresholds; i.e. $1/k^2-m^2 +i\varepsilon$ in propagating 
states.  However, at finite $\lambda$, where one is effectively 
extending the analysis on the field theory side to an infinite 
number of Feynman diagrams, this is in accord with extracting 
unitarity cuts in two-particle channels of multi-loop Feynman diagrams 
of arbitrary loop order.  In each order within a loop expansion in 
perturbation theory of the 
correlation function, one expects a contribution of multi-particle 
cuts in the two-particle channel (for example, the double-box Feynman 
diagram contributes both two- and three-particle cuts).  This 
holographic unitarity feature appears to accord with 
expectations from field theory although the positivity condition is 
not clear in field theory.

This factorization of the 
holographic $S$-matrix elements at tree-level was noted in \cite{cs2}, and 
follows from the fact that there is no multi-body phase space integration 
to be carried out at string tree-level on the anti-de Sitter space:  
it is a property of the classical string description  
of the large $N$ limit of the gauge theory.  
Unitarity restricts the functions $M_i$ to be real.  Holding 
$k_1=k_4$ and $k_2=k_3$ (and other quantum numbers associated to these 
lines the same) we sum over all terms contributing  
on the right-hand side of the sum; each term contributes a 
square of a real function and thus must be positive (strictly, 
non-negative).  However, 
we must be concerned with the presence of an infinite number of 
terms.   For any finite number of terms the sum must be positive for 
any finite (positive) value of $s$.  If the sum becomes negative after 
taking into account the infinite number of terms, then a pole or cut is 
evidenced, reflecting a finite radius of convergence.     

The non-negativity condition does not depend on the detailed form 
of string perturbation 
theory in a Ramond-Ramond background, but rather on the use of 
field equations in a Ramond-Ramond background.  We should note 
that bulk four-point vertices contribute to the imaginary parts at 
exceptional values of momenta for the massless fields, as at 
$s=0$, but this limit does not enter into this analysis. 

We now compare with the field theory predictions arising 
at the free limit, i.e. $\lambda=0$.  We normalize the protected 
gauge invariant operators, ${\rm Tr} \phi^{(i_i} \ldots \phi^{i_k)}(z)$, 
where the SO(6) vectors $\phi^i$ are the $N=4$ scalars in the 
adjoint representation of SU(N), in the form, 
\bq  
{\cal O}(z)= {N\over (g^2N)^{k/2}} {\rm Tr}~ \phi^{(i_i} 
\ldots \phi^{i_k)}(z) \ , 
\fq 
so that the free-field limit is independent of $\lambda$ and 
proportional to $N^2$; the Lagrangian has the microscopic 
fields suitably scaled to agree with the factor $N^2$ of correlators 
of the gauge invariant composite operators in the free-field 
approximation.  We are only considering the symmetrized 
traceless tensor product of scalar fields $\phi^{i}(z)$ for 
simplicity.   

In this normalization, the free-field result for the four-point 
function of symmetric bi-linear operators $O(z) = {1\over g^2} 
{\rm Tr}\phi^{(i} \phi^{j)}(z)$ is explicitly in 
$x$-space the box diagram and is proportional to $N^2$.  In four 
dimensions, we have for the (unrenormalized) result, where the 
propagator is $\Delta(z) ={1\over 4\pi^2} {1\over z^2-i\varepsilon}$,  
\bq 
<\prod_j {\cal O}^{m_j n_j} (z_j)> = {N^2\over (4\pi^2)^4} 
 T^{\{mn\}} \Bigl[ {1\over (z_{12}^2 -i\varepsilon) 
(z_{23}^2-i\varepsilon) ( z_{34}^2-i\varepsilon) ( z_{41}^2-i\varepsilon) } 
\fq 
\bq 
+ 
{1\over (z_{12}^2 -i\varepsilon)(z_{13}^2 -i\varepsilon) 
 (z_{34}^2 -i\varepsilon) (z_{42}^2 -i\varepsilon)} 
\fq 
\bq 
+  {1\over (z_{13}^2 -i\varepsilon) (z_{23}^2 -i\varepsilon) 
 (z_{24}^2 -i\varepsilon) (z_{41}^2 -i\varepsilon)} \Bigr] \ ,
\label{xspacebox}
\fq 
where $T^{\{mn\}}$ is the group theory factor associated with the 
single trace-operators and $z_{ij}^2=(z_i-z_j)^2$.  Similar results may be 
obtained regarding other free-field correlators in $x$-space.

We choose to obtain the result in $k$-space, where we may compare 
with previous results.  After taking the Fourier transform, and for external 
momenta satisfying $k_j^2=0$, we obtain the following expression for the 
box diagram \cite{integrals}, with ordering of momenta at the vertices 
$k_1\ldots k_4$, 
\bq  
I(s,t) = N^2 {c(\epsilon)\over st} \Bigl[ {1\over \epsilon^2} \bigl( 
(-s)^{-\epsilon} + (-t)^{-\epsilon} \bigr) -{1\over 2} \ln^2 (-s/-t) 
 \Bigr] \ . 
\label{kspacebox}
\fq 
Here, 
\bq  
I(s,t) = -i\int {d^dl\over (2\pi)^d} {1\over l^2 (l-k_1)^2 (l-k_1-k_2)^2 
 (l+k_4)^2 }  \ , 
\label{mombox}
\fq 
where, in including the appropriate $i\varepsilon$, we have to 
change $s_{ij} \rightarrow s_{ij} + i\varepsilon$.  The normalization 
in \rf{mombox} explicitly has a factor of $-i$ to account for the 
removal of $i$ in the dual string theory calculation \rf{ref}.  
The dimension dependent constant $c(\epsilon)$ is   
\bq  
c(\epsilon) = {1\over 2(4\pi)^2} {\Gamma(1+\epsilon)\Gamma^2(1-\epsilon) 
\over \Gamma(1-2\epsilon)} \ . 
\fq 
Note that by 
analytic continuation from higher than four dimensions the box 
diagram evaluated at zero momentum is formally zero, i.e. 
$I^\epsilon(k_i=0)=0$.  We have also regulated the above integral 
using dimensional reduction, where $d=4-2\epsilon$; 
holding $\epsilon<0$ regulates the infra-red divergences appearing 
within the above integration.  The integral is ultra-violet finite 
but keeping $\epsilon>0$ would have regulated this occurence in 
any case.  The complete expression for the four-point correlator in 
the free-field limit of bi-linear operators is the permuted sum 
of the integrals, i.e. $I(s,t)+I(s,u)+I(t,u)$.  

The expansion in $\epsilon$ of the particular (infra-red divergent) 
integral in \rf{kspacebox} is explicitly 
\bq  
I(s,t) = N^2 {c(\epsilon)\over st} \Bigl[ {1\over \epsilon^2} 
 - {1\over\epsilon} \ln(st) - {1\over 2} \ln(-s) \ln(-t)  
 \Bigr] \ . 
\label{kspaceexpand}
\fq 
The imaginary parts in the two-particle channels may be  
extracted from the expanded form in \rf{kspaceexpand} via 
complex conjugation.  

To end our discussion of the box 
integrations we give the most general case, which would hold 
for non-exceptional values of external momenta.    
The result for general momenta $k_j^2\leq 0$ is more 
complicated, but may be written down in terms of known functions 
as \cite{integrals}
\bq  
I^{4m} = {1\over r (st-k_1^2 k_2^2)} \Bigl[ 
 {\rm Li}_2\bigl[{1\over 2}(1+r)\bigr] - {\rm Li}_2\bigl[{1\over 2}(1-r)\bigr] 
\fq 
\bq  
 + {\rm Li}_2\bigl[-{1\over 2\lambda}(1-2\lambda-r)\bigr] 
 - {\rm Li}_2\bigl[-{1\over 2\lambda}(1-2\lambda+r)\bigr]
- {1\over 2}\ln(\lambda) \ln\bigl({1+r\over 1-r}\bigr) \ ,   
\fq 
where 
\bq  
r=[1-4\lambda]^{1/2} ,\qquad\qquad {d\over dx} {\rm Li}_2(1-x) = 
 {\ln x\over 1-x} \ , 
\fq 
and, 
\bq  
\lambda=-{k_1^2 k_3^2\over st} \ . 
\fq 
This box integral becomes infra-red divergent when we take any 
$k_j^2\rightarrow 0$: soft for individual null momenta, and collinear 
for any two momenta flowing into adjacent vertices becoming null.  
If we maintain the dimensionally regularized form, keeping the full 
$\epsilon$ dependence of the above (not shown), then the preceeding 
box integral follows.  

We now find the imaginary parts in the correlation functions 
in the free-field limit arising from the three contributing permuted 
box diagrams.  Holding $s>0$ and $t,u<0$, together with 
$k_j^2\leq 0$, we obtain for the imaginary part, 
\bq  
{\rm Im}_s I(s,t) = -N^2 \pi {c(\epsilon)\over st} \Bigl[ {1\over \epsilon} 
+ {1\over 2} \ln \vert t\vert \Bigr] \ , 
\fq 
together with the $t\leftrightarrow u$ contribution, 
\bq 
{\rm Im}_s I(s,u) = -N^2 \pi {c(\epsilon)\over su} \Bigl[ {1\over \epsilon} 
+ {1\over 2} \ln \vert u\vert \Bigr] \ .
\fq 
The net result for the imaginary part is then: 
\bq  
{\rm Im}_s \Bigl[ I(s,t) + I(s,u) + I(u,t) \Bigr] = 
 N^2 \pi c(\epsilon) \Bigl[ {1\over \epsilon} {1\over tu} 
 - {1\over 2 st} \ln\vert t\vert - {1\over 2 su} \ln\vert u\vert \Bigr] 
\ .
\fq 
This simple analysis shows that, depending on whether $\vert t\vert$ and 
$\vert u\vert$ is greater or lesser than one, we may change the sign 
of the imaginary part.  The value of $\epsilon$ also enters into the 
above infra-red divergent integral and may change its sign; we shall 
take into account the interchange of limits as we examine the unitarity 
relation between weak and strong 't Hooft coupling.  

As we take $k_1=k_4$ and $k_2=k_3$, we force the Mandelstam invariants 
to the limit $t=0$ and $-u=s=x$.  The limiting form of the imaginary 
part in this case is 
\bq 
{\rm Im}_s \Bigl[ I(s,t) + I(s,u) + I(u,t) \Bigr] = 
- N^2 \pi c(\epsilon) {1\over tx} \Bigl[ {1\over\epsilon} + 
{1\over 2} \ln\vert t\vert 
 \Bigr] + N^2 \pi c(\epsilon) {1\over 2x^2} \ln\vert x\vert \ , 
\fq 
and is apparently negative for fixed $x$ and $t\rightarrow 0^-$ for 
fixed $\epsilon$ very small.  For $t$ infinitesimally small, the value 
is positive or negative depending on the value of $\epsilon$. 
The full dimensionally regularized expression, that is the result 
for non-zero $\epsilon$ at $t=0$, is non-vanishing but does not 
possess an imaginary part (For example, the integral $I(t,u)$ reduces 
to $(-u)^{-2-\epsilon}$ times a constant and is purely 
real.  Remaining integrals are similarly evaluated.)  

The preceeding calculation in the tree-level IIB superstring theory on 
$AdS_5\times S^5$ in a Ramond-Ramond background always gave a positive 
value.  In the case of external string states in our amplitude 
calculation only internal propagating string modes contribute, 
but the analysis may be performed for general four-point correlation 
functions.  The calculation of the imaginary parts arising from the string 
modes in the AdS space translates into evaluating the large $N$, 
finite $\lambda$, imaginary part of the holographic $S$-matrix element.   

A conundrum arises when we compare between the large $\lambda$ and 
small $\lambda$ regimes.  From explicitly evaluating the two different 
integrals we conclude that the imaginary part must pass through zero, 
or becomes zero at the value $\lambda=0$ within dimensional regularization.  
Define this critical value of the coupling as $\lambda_0(k_j)$.  
At a finite value of the coupling, such as $\lambda_0$, only the 
positive functions, the squares in \rf{factcond}, 
contribute.  This mismatch indicates a finite radius of convergence 
of the perturbative series defined either by the $N=4$ field theory at 
large $N$, or at holographic string tree-level.  Hence naively the two 
regions do not agree with eachother.\footnote{It is possible 
that this comparison between the field theory and string theory is 
regularization dependent, but the string theory obeys the 
factorization in \rf{factcond} also in the radial coordinate 
regulating scheme $z_0>r_c$ in addition with dimensional 
reduction.  The two types of delta function terms, those proportional 
to finite or infinite coefficients \cite{cs3}, are not encountered 
directly here because we are working in momentum space and have 
not examined possible AdS boundary term effects.}  

There are several options to account for this disagreement: (1) There 
is a phase transition at finite $\lambda$ and at large $N$ in 
the dual theory to IIB superstring theory reflecting the divergence 
of the tree-level holographic scattering element (related to 
\cite{gw}); (2) Related is that the series at infinite $N$ has a 
finite radius of convergence in $1/\lambda$ from $\lambda$ large: there 
is either a pole on the positive real axis of $\lambda$ or a cut passing 
through or laying on top of it (possibly also in a Borel transform 
reflecting a finite radius of convergence); (3) Unitarity is not 
preserved in the holographic $AdS$ formulation; (4) There are 
modifications yet to be specified in the matching between $N=4$ 
super Yang-Mills theory and IIB superstring theory on $AdS_5\times S^5$ 
at finite $\lambda$, away from $\lambda=\infty$.  

Instanton-like effects may be predicted in 
the case of option (1) or (2) similar to the $e^{-1/g_s}$ effects predicted 
in the analysis of large orders of perturbative string theory to 
restore convergence; likewise, within the correspondence we may 
infer similar contributions missed in $N=4$ super Yang-Mills 
theory which, unlike the 't Hooft instanton, contribute in the 
infinite $N$ limit and are not easily seen with the use of 
field equations. (It is interesting that 
in unbroken $N=4$ super Yang-Mills theory the BPS mass formula indicates 
an infinite number of massless dyons in the theory.)  The existence of 
further contributions is supported by the evidence for finite radii of 
convergence in perturbative expansions in the large $N$ limit \cite{thooft} 
and evidence for potential non-Borel summability in various field 
theories.  If present, these effects would deserve dual $AdS$ 
descriptions possibly 
via the inclusion of additional states or processes within the 
string theory side (and contributing at intermediate coupling 
in $\lambda$).  Similar behavior is found in the context of 
perturbative string 
theory \cite{gp,ss} and transitions have been argued for in 
$N=4$ super Yang-Mills theory at finite temperature \cite{ml}.  
Additional modifications would have to leave the two- and three-point 
functions of chiral primary operators unchanged at leading order 
in $N$ to maintain the validity of previous work regarding the 
matching \cite{fmmr,cs1,lmrs,dfs}.  

Another option is that the higher-order terms in the $\alpha'$ 
expansion of the non-abelian Dirac-Born-Infeld 
action are to be included in the super Yang-Mills theory description.  
This option would suggest that $N=4$ super Yang-Mills theory is 
only dual to IIB string theory on AdS at $\lambda=\infty$.  Sub-leading 
effects would necessitate these higher order terms in the DBI 
expansion.  However, this possibility is problematic in that 
the DBI action is non-renormalizable in four dimensions and further 
inclusion of terms might alter the two- and three-point correlation 
functions at sub-leading order.  

While this work was being completed, an analysis of the effects of  
finite radii of convergence in an AdS/CFT correspondence have 
been presented in the context of non-supersymmetric type 0B 
theory \cite{ik}.

\vskip .3in 
\noindent Acknowledgements 
\vskip .2in 

G.C. would like to thank Zvi Bern, Miao Li, Emil Martinec, Shiraz 
Minwalla, Koenraad Schalm, Warren Siegel and Herman Verlinde for 
useful discussions and comments.  This work is supported in part by 
the U.S. Department of Energy, Division of High Energy Physics, 
Contract W-31-109-ENG-38.  The hospitality and stimulating 
environment of the Cargese 99 Summer institute is acknowledged 
during which time much of this work was done.

\end{document}